\documentclass{IEEEtran}

\usepackage{pifont,epsfig,amsmath,amssymb,latexsym,stmaryrd}
\usepackage[all]{xy}

\newdir{pb}{:(1,-1)@^{|-}}
\def\pb#1{\save[]+<16 pt,0 pt>:a(#1)\ar@{pb{}}[]\restore}

\usepackage{url}

\usepackage{tikz}
\usetikzlibrary{calc,positioning}
\usepackage{graphicx}
\usepackage{tikz}
\usepackage{amsmath}
\usepackage{amssymb}
\usepackage{cmll}
\usepackage{amsthm}
\usepackage{stmaryrd}
\usepackage{MnSymbol}
\usepackage{mathtools}

\newcommand{\hide}[1]{}
\newcommand{\drp}[2]{d_v[{#1}; #2]}
\newcommand{\ket}[1]{{|{{#1}}\rangle}}
\newcommand{\Tr}{\mathop{{\rm Tr}}}

\newcommand{\sym}[1]{\widetilde{#1}}
\newcommand{\ssym}[1]{\widetilde{\widetilde{#1}}}

\newcommand{\Sup}{{\rm Sup}}
\newcommand{\Inf}{{\rm Inf}}

\newcommand{\pfrule}[2]{{{\hbox{{$#1$}}\over{\strut \hbox{{$#2$}}}}}}
 \def\fatbar{\mathop{[\!]}}

\newcommand{\tri}{\unlhd}
\newcommand{\sscirc}{\oast}

\newcommand{\lam}{\lambda}

\newcommand{\stcomp}[3]{\exists #1.\, [\, #2 \mathrel{
\vvbar} #3\, ]}

\newcommand{\Eta}{{\rm H}}
\newcommand{\Ga}{\Gamma}

\theoremstyle{remark}

\newtheorem*{theorem*}{Theorem}

 \usepackage[all]{xy}

\newcommand{\opmove}{\ominus}
\newcommand{\plmove}{\oplus}
 
\def\profto{\!\!\!\xymatrix@C-.75pc{\ar[r]|-{\! +\!} &}\!\!\! }

\newdir{pb}{:(1,-1)@^{|-}}
\def\pb#1{\save[]+<16 pt,0 pt>:a(#1)\ar@{pb{}}[]\restore}

\newcommand{\vvbar}{{\mathbin{\parallel}}}
\newcommand{\scirc}{{{\odot}}}

\newcommand{\longcov}[1]{{\stackrel{#1}{\mathrel-\joinrel\relbar\joinrel\subset\,}}}

\newcommand{\imc}{\rightarrowtriangle}

\newcommand{\sig}{\sigma}

\newcommand{\CC}{{\rm C\!\!C}}
\newcommand{\cc}{\gamma}

\newcommand{\bel}{\sqsubseteq}
  \newcommand{\spanpl}[5]{{\xymatrix{
    & {#3}\ar@{_{(}->}[dl]_{#2}\ar[dr]^-{#4} &\\
      {#1} && {#5} 
  }}}

\def\Con{{\rm Con}}

\newcommand{\esws}{event structure with symmetry}
\newcommand{\essws}{event structures with symmetry}
\newcommand{\eswp}{event structure with polarity}

\newcommand{\iconf}[1]{\:\!{\cal C}^\infty(#1)}

\newcommand{\arr}[1]{{{\stackrel{#1}{\longrightarrow}}}}
\newcommand{\id}{{\rm id}}

\newcommand{\set}[2]{{\{  #1\  | \  #2 \} }}
\newcommand{\setof}[1]{{\{ #1 \} }}

\newcommand{\eqdef}{\mathrel{=_{\mathrm{def}}}}

\newcommand{\iso}{\cong}

\def\mxxth{\mathsurround=0pt}
\dimendef\dimenxx=0
\def\openup{\afterassignment\xxpenup\dimenxx=}
\def\xxpenup{\advance\lineskip\dimenxx
  \advance\baselineskip\dimenxx \advance\lineskiplimit\dimenxx}
\def\eqalign#1{\,\vcenter{\openup1\jot \mxxth
  \ialign{\strut\hfil$\displaystyle{##}$&$\displaystyle{{}##}$\hfil
     \crcr#1\crcr}}\,}

\newif\ifdtxxp
\def\displxxy{\global\dtxxptrue \openup1\jot \mxxth
  \everycr{\noalign{\ifdtxxp \global\dtxxpfalse
      \vskip-\lineskiplimit \vskip\normallineskiplimit
      \else \penalty\interdisplaylinepenalty \fi}}}
\def\displaylines#1{\displxxy
  \halign{\hbox to\displaywidth{$\hfil\displaystyle##\hfil$}\crcr
      #1\crcr}}

\newskip\mycntring \mycntring=0pt plus 1000pt minus 1000pt

\def\leqalignno#1{\displxxy \tabskip=\mycntring
  \halign to\displaywidth{\hfil$\displaystyle{##}$\tabskip=0pt
      &$\displaystyle{{}##}$\hfil\tabskip=\mycntring
      &\kern-\displaywidth\rlap{$##$}\tabskip=\displaywidth\crcr
      #1\crcr}}

\newcommand{\ie}{{\it i.e.}}
\newcommand{\eg}{{\it e.g.}}

\newcommand{\viz}{{\it viz.}}

\newdir{C}{%
!/4.5pt/@{( }*:(1,-.2)@{}*:(1,+.2)@_{}}

\newdir{|>}{%
!/4.5pt/@{|}*:(1,-.2)@{>}*:(1,+.2)@_{>}}

\begin{document}

\title{
Distributed Games and Strategies 
}
\author{Glynn Winskel,  Computer Laboratory, University of Cambridge}

\maketitle



\begin{abstract}
A summary of work on distributed games and strategies done within the first three years of the ERC project ECSYM is presented.
\end{abstract}

\section{Motivation}
\newcommand{\vs}{{\it vs.}}
My current interest is to develop a mathematical theory of 
 {\em distributed} games in which a Player (or a team of players) can interact and compete against an Opponent (or a team of opponents) in a highly distributed fashion, without, for instance, enforcing that their moves occur in a sequential fashion, or need to alternate.  Although phrased in terms of 
 `Player' and `Opponent'    the dichotomy Player \vs~Opponent can stand for 
Process \vs~Environment,  Prover \vs~Disprover,  or 
Ally \vs~Enemy. 
These alternatives indicate the wide range of potential applications in computer science, logic,  and beyond.

 The original motivation for developing distributed games was as a framework 
   to represent and analyse the behaviour of computation, which can be distributed and highly interactive.   It had become clear that many aspects of computation did not fit within traditional approaches.  There was a need 
to tackle anomalies such as non-deterministic dataflow, which lay outside the usual techniques, and repair the divide between the two main approaches of denotational and operational semantics.  Roughly,
in representing computation, the role of Player is to represent that part of the system which is controllable and could be programmed, while that of Opponent stands for the unforeseeable and uncontrollable environment. 

Strategies are as potentially fundamental as relations and functions. It is surely because of our limited mental capacity, and not because of its unimportance, that the mathematical concept of strategy has been uncovered relatively late. It is hard to think about the successive contingencies involved in playing a game in the same way that is hard to think about interacting processes. Developing strategies in the extra generality demanded by distributed interaction has enabled us to harness computer-science expertise in structure and distributed  computation in their understanding and formalisation.
The extra generality has  
 revealed more clearly their  essential mathematical nature,  
new insights, and a mathematical robustness to the concept of strategy. 

As the work has developed it has become increasingly clear that
strategies are at the crossroads of several areas both semantic and algorithmic,  and beyond computer science and logic,  in other sciences, where perhaps the  structural emphasis, so part and parcel to computer science, has been missing.  In this note I have tried to summarise the work on distributed games and strategies done 
over the last three years, and indicate how it is to be developed 
across a range of areas. \\


\noindent
{\bf Outline}
The presentation begins with an introduction to event structures (Section~\ref{sec:evstrs}
), 
the fundamental model on which we base the behaviour of distributed games; plays in games will take the form of partial orders of moves.  We then 
 develop the definition of distributed strategies (Section~\ref{sec:DistgamesandStrats}), adopting as the guiding principle that certain `copy-cat' strategies should behave as identities w.r.t.~a composition of strategies; the definition of composition 
is driven by two fundamental operations on games, those of parallel composition and dual.
  Distributed games and strategies have a very rich structure, leading to  a language for programming strategies (Section~\ref{sec:language}).   The language is robust w.r.t.~extensions by winning conditions (Section~\ref{sec:winconds}), to probabilistic strategies (Section~\ref{sec:prob}), quantum strategies (Section~\ref{sec:quan}), games with payoff functions (Section~\ref{sec:payoff}) and of imperfect information (Section~\ref{sec:impinfo}) and games with symmetry (Section~\ref{sec:sym}). 
We conclude with remarks on extensions, present limitations, and ambitions for the future (Sections~\ref{sec:extnsandlims} and~\ref{sec:future}).

\section{Event structures~\cite{evstrs}}\label{sec:evstrs}

The behaviour of distributed games  is based on event structures, rather than trees.  Instead of regarding a play in a game as a sequence of Player and Opponent moves it is given the structure of a partial order of occurrences of moves.  

Event structures describe a process, or system, in terms of its possible event occurrences, their causal dependency and consistency. Just as it can be helpful to  
understand the behaviour of a state-transition diagram in terms of its unfolding to
  a tree,  more detailed models, such as Petri nets, which make explicit the local nature of events and their changes on state,  unfold to an event structure~\cite{NPW}.   In this sense event structures are a concurrent, or distributed, analogue of trees; 
though in an event structure the individual `branches' are no longer necessarily sequences but have the shape of a partial order of events. 

An {\em event structure}  comprises $(E, \leq, \Con)$,   consisting of a set $E$ of {\em events} (really event occurrences) which are
partially ordered by $\leq$, the {\em causal dependency
relation},
and a  nonempty {\em consistency relation} $\Con$ consisting of finite subsets of $E$.  The relation
$e'\leq e$ expresses that event $e$ causally depends on the previous occurrence of event $e'$.  That a finite subset of events is consistent conveys that its events can occur together by some stage in the evolution of the process.  
Together the relations satisfy several  axioms:
$$
\eqalign{
&\set{e'}{e'\leq e}\hbox{ is finite for all } e\in E,\cr
&\setof{e}\in\Con \hbox{  for all } e\in E,\cr
&Y\subseteq X\in\Con \hbox{ implies }Y\in \Con,\ \hbox{ and}\cr
&X\in\Con \ \&\  e\leq e'\in X \hbox{ implies } 
X\cup\setof{e}\in\Con.\cr}
$$
The first axiom says that an event causally depends on only a finite number of events, the second that there are no redundant events, which are in themselves inconsistent.  The third axiom expresses the reasonable property that a subset of consistent events is consistent, while the final axiom entails that the $\leq$-down-closure of any consistent set of events is also consistent.  
Two events $e, e'$ are considered to be {\em concurrent} if  the set $\setof{e,e'}$ is in $\Con$ and neither event is causally dependent on the other.   

It is sometimes convenient to draw event structures.  For example, 
$$\small
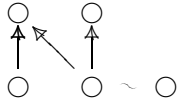
\xymatrix@R=14pt@C=13pt{
\bigcirc &\bigcirc&\\
\bigcirc\ar@{|>}[u]\ar@{|>}[u]&\bigcirc\ar@{|>}[ul]\ar@{|>}[u]\ar@{~}[r]& \bigcirc
}
$$ 
illustrates an event structure consisting of five events where, in particular,  the top event on the left causally depends on the previous occurrences of two concurrent events --- the arrows express the causal dependency --- one of which is inconsistent with the event on the far right --- we have indicated the inconsistency between the two events by a wiggly line.

Given this understanding of an event structure, there is an accompanying notion of state, or history, those events that may occur  up to some stage in the behaviour of the process described.  A {\em configuration} is a, possibly infinite, set of events $x\subseteq E$ which is both  consistent and down-closed  w.r.t.~causal dependency:
\begin{itemize} 
\item[]
{\em Consistent:} $X\subseteq x  \hbox{ and }  X \hbox{ is finite}  \hbox{ implies } X\in\Con$\,,
and
\item[]
{\em Down-closed:}
$ e'\leq e\in x  \hbox{ implies } e' \in x$.
\end{itemize} 
An individual configuration inherits a partial order from the ambient event structure, and represents one possible  partial-order history.  

It will be very useful to relate event structures by maps.  
 A {\em map} of event structures 
$f:E
\to E'$
 is a partial function  
$f$ from $E$ to $E'$ such that
the image of a configuration $x$ is a configuration $f x$ and any event of $f x$  arises as the image of a unique event of $x$.  
 In particular, when $f$ is a total function it restricts to a bijection $x\iso f x$ between any configuration $x$ and its image $f x$.  

A map $f:E\to E'$ preserves concurrency:  if two events in $E$ are concurrent, then their images if defined are also concurrent. The map also reflects causal dependency locally, in the sense that if $e, e'$ are events in a configuration $x$ of $E$ for which $f(e')\leq f(e)$ in $E'$, then $e'\leq e$ also in $E$; the event structure $E$ inherits causal dependencies from the event structure $E'$ via the map $f$.  In general a map of event structures need not preserve causal dependency; when it does we say it is {\em rigid}. 

In describing distributed games and strategies we shall rely on two properties of maps.  Firstly, any map of event structures $f:E\to E'$ factors into the composition of a partial map of event structures followed by a total map of event structures
$$
E\arr{p} E_0 \arr{t} E'
$$
in such a way that for any other factorisation 
$
E\arr{p_1} E_1 \arr{t_1} E'
$
with $p_1$ partial and $t_1$ total, there is a unique (necessarily total) map $h:E_0\to E_1$ such that 
$$
\xymatrix@R=12pt@C=16pt{&E_1\ar[dr]^{t_1}&\\
E\ar[ur]^{p_1}\ar[r]^{p} &E_0 \ar@{..>}[u]_h\ar[r]^{t} &E'}
$$
commutes.  
The event structure $E_0$ is obtained as the ``projection'' of the event structure $E$ to the events on which $f$ is defined.  We call the total map $t$ {\em the defined part} of $f$.

Secondly we shall use pullbacks of total maps.  Pullbacks are an important construction in representing a process built from two processes sharing a common interface.    
Maps   $f:A\to C$ and $g:B\to C$ always have pullbacks in the category of event structures, but they are more simple to describe in the case where $f$ and $g$ are total, and this is all we shall need:
$$
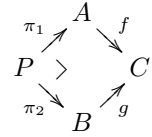
\xymatrix@R=8pt@C=8pt{
&A\ar[rd]^{f}&\\
P\pb{0}\ar[ur]^{\pi_1}\ar[rd]_{\pi_2}&&C\\
&B\ar[ur]_{g}&}
$$
Roughly, configurations of the pullback $P$ are matches between configurations of $A$ and $B$ which satisfy the causal constraints of both.  Precisely, 
 finite configurations of $P$ correspond 
to composite bijections $$\theta: x\iso fx = gy\iso y$$ between finite configurations $x$ of $A$ and $y$ of $B$ such that $f x= g y$,
for which the transitive relation generated on $\theta$ by  $(a,b)\leq (a',b')$ if $a\leq_A a'$ or $b\leq_B b'$ 
has no non-trivial causal loops, and so forms a partial order.

\section{Distributed games and strategies --- the definitions~\cite{lics11}}\label{sec:DistgamesandStrats}
Often the behaviour of a game is represented by a tree in which the arcs correspond to occurrences of moves by Player or Opponent.  Instead we can represent the behaviour of a distributed game more accurately by an event structure together with a polarity function from its events to + or $-$ to signify whether they are move occurrences of Player or Opponent, respectively.

A game might generally have winning conditions, a subset of configurations at which Player is deemed to have won, or more generally a payoff function from configurations to 
the reals.  

There are two fundamentally important operations on two-party games.  One is that of forming the dual game in which the moves of Player and Opponent are reversed.  On an event structure with polarity $A$ this amounts to reversing the polarities of events to produce the dual $A^\perp$.  By a strategy in a game we will mean a strategy for Player.  A strategy for Opponent,  or a counter-strategy,  in  a game $A$ will be identified with a strategy in $A^\perp$.
The other operation is a simple parallel composition of games, achieved on event structures with polarity $A$ and $B$ very directly by simply juxtaposing them, ensuring a finite subset of events is consistent if its overlaps with the two games are individually consistent, to form $A\vvbar B$.   

As an example of a strategy in a game consider the {\em copy-cat} strategy for a game $A$.  This is a strategy in the game $A^\perp\vvbar A$ which, following the spirit of a copy-cat, has   Player moves  copy the corresponding Opponent moves in the other component.  In more detail, the copy-cat strategy  $\CC_A$ is obtained by adding extra causal dependencies to $A^\perp\vvbar A$ so that any Player move in either component  causally depends on its copy, an Opponent move, in the other component.  It can be   checked that this generates a partial order of causal dependency.  A finite set is taken to be consistent if its down-closure w.r.t.~the order generated is consistent in $A^\perp\vvbar A$.   We illustrate the construction on the simple game comprising a Player move causally dependent on a single Opponent move: 
$$
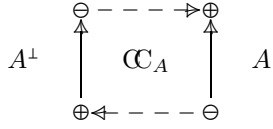
\xymatrix@R=6pt@C=6pt{
&\opmove \ar@{--|>}[rr]&& \plmove& \\\
A^\perp&&\CC_A&&A\\
&\plmove\ar@{|>}[uu] && \opmove\ar@{|>}[uu] \ar@{--|>}[ll]&
}
$$

In characterising the configurations of the copy-cat strategy an important partial order on configurations is revealed.  Clearly configurations of a game $A$ are ordered by inclusion $\subseteq$.  For configurations $x$ and $y$, write $x\subseteq^- y$ and $x\subseteq^+ y$ when all the additional events of the inclusion are purely Opponent, respectively,  Player moves.
A configuration $x$  of $\CC_A$ is also a configuration of $A^\perp\vvbar A$ and as such splits into two configurations  $x_1$ on the left and $x_2$ on the right. The extra causal constraints of copy-cat ensure that the configurations of $\CC_A$ are precisely those configurations of $A^\perp\vvbar A$ for which 
it holds that 
$$ 
 x_2\bel_A x_1 \,, \hbox{ defined as }
%
x_2 \supseteq^- y\subseteq^+ x_1\,,
$$
 for some configuration $y$.  The relation $\bel_A$ is in fact a partial order on configurations.   Increasing in the order $\bel_A$ involves losing Opponent moves and gaining Player moves.  Because it generalises 
the pointwise order of domain theory, initiated by Dana Scott, we call the order $\bel_A$ the {\em Scott order}.  

Strategies in a game are not always obtained by simply adding extra causal dependencies to the game. For example,  consider the game comprising two Opponent moves in parallel with a Player move
and the strategy (for Player) in which Player make their move if Opponent makes one of theirs.  Here the strategy is represented by
$$
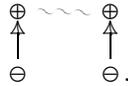
\xymatrix@R=12pt@C=8pt{\plmove\ar@{~}[rr]&&\plmove\\
\opmove\ar@{|>}[u]&& \opmove\,.\!\!\!\ar@{|>}[u]}
$$
We are forced to split the Player move of the game into two moves, each dependent on different Opponent moves, and mutually inconsistent indicated by the wiggly line.  For reasons such as this we are led to separate the actual moves of the strategy into an \eswp~$S$ and, in order to track how actual moves correspond to moves in the game,
formalise a strategy in a game $A$ as a total map of event structures 
$$
\sig:S\to A
$$
which preserves polarity.  (We have met a very special case of this in the copy-cat strategy where the role of $S$ is taken by $\CC_A$ and $\sig$ acts as the identity function on events.)
The event structure $S$ describes the possibly nondeterministic plays of the strategy. Automatically a state of play of the strategy, represented by a configuration  $x$ of $S$, determines a position of the game,  a configuration $\sig x$ of $A$.  Directly from the fact that $\sig$ is a map, we know that any move in $\sig x$ is due to the play of a unique move in $x$.  The total map $\sig:S\to A$ really just expresses that $S$ represents a nondeterministic play in the game $A$.  More is expected of a strategy.  For example, consider the game consisting of a Player move concurrent with a move of Opponent and the two total maps indicated: 
\vskip 0.05in
 
$
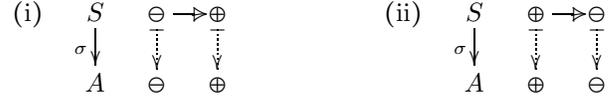
\xymatrix@R=12pt@C=10pt{
{\rm (i)}& S\ar[d]_\sig& \opmove
  \ar@{|>}[r]
 \ar@{|.{>}}[d]
 &\plmove\ar@{|.{>}}[d]\\
&A& \opmove&\plmove
 }
 $
 \qquad\qquad\quad
 $\xymatrix@R=12pt@C=10pt{
{\rm (ii)} &S\ar[d]_\sig& \plmove
  \ar@{|>}[r]
 \ar@{|.{>}}[d]
 &\opmove\ar@{|.{>}}[d]\\
&A& \plmove&\opmove
 }
 $\\

\noindent
The first map (i) seems reasonable as a strategy; Player awaits the move of Opponent and then makes a move. However, the second map (ii) seems dubious;  Player forces Opponent to wait until they have made their move, inappropriate in a distributed strategy.  

Instead of guessing, we seek a principled way to determine what further properties a strategy should satisfy.  
In fact, the further conditions we shall impose on strategies will be precisely those needed to ensure that the copy-cat strategy behaves as an identity w.r.t.~the composition of strategies.\footnote{We 
consider two strategies $\sig:S\to A$ and $\sig':S'\to A$ to be essentially the same 
if there is an isomorphism $f:S\iso S'$ of event structures respecting polarity such that $\sig = \sig'f$.}    
To do so we 
adapt an important idea of Conway followed up by Joyal, explaining how to extend the notion of strategy {\em in} a game to that of a strategy {\em between} games~\cite{conway, joyal}.  The operations of dual and simple parallel composition of games are the key.  
  
A  strategy  {\em from} a game $A$ {\em to} a game $B$ is a strategy in the compound game $A^\perp \vvbar B$.  In particular, copy-cat of a game $A$ is now seen as a strategy from $A$ to $A$. 

 In composing two strategies one $\sig$ in $A^\perp\vvbar B$ and another $\tau$ in $B^\perp\vvbar C$  
one firstly instantiates the Opponent moves in component $B$ by Player moves in $B^\perp$ and {\it vice versa}, and then secondly hides the resulting internal moves over $B$.  The first step is achieved efficiently via pullback.
Temporarily ignoring polarities, the pullback
  
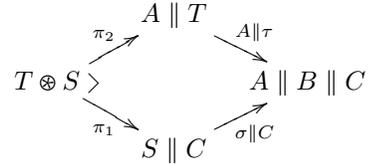
\[
\xymatrix@R=10pt@C=10pt{
&A\parallel T	\ar[dr]^{A\parallel \tau}\\
T\sscirc S	\ \ 	\ar[ur]^{\pi_2}
		\ar[dr]_{\pi_1}
		\ar@{}[rr]|{ \ \ }
		\pb{0}&&
A\parallel B \parallel C\\
&S\parallel C	\ar[ur]_{\sigma \parallel C}
}
\]
 ``synchronises'' matching moves of $S$ and $T$ over the game $B$.  But we require a strategy over the game $A^\perp\vvbar C$ and the pullback $T\sscirc S$ has internal moves over the game $B$.  We achieve this via the projection  of  $T\sscirc S$ to its moves over $A$ and $C$.  We make use of the partial map from $A\vvbar B\vvbar C$ to $A\vvbar C$ which acts as the identity function on $A$ and $C$ and is undefined on $B$. 
The composite partial map
 
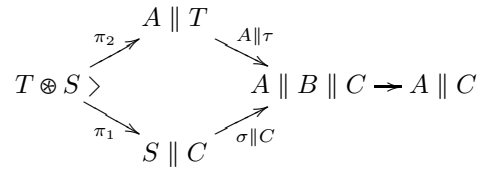
\[
\xymatrix@R=10pt@C=10pt{
&A\parallel T	\ar[dr]^{A\parallel \tau}\\
T\sscirc S\ 	\ 			\ar[ur]^{\pi_2}
		\ar[dr]_{\pi_1}
		\ar@{}[rr]|{ \ \ }
		\pb{0}&&
A\parallel B \parallel C
		\ar[r]&
A\parallel C\\
&S\parallel C	\ar[ur]_{\sigma \parallel C}
}
\]
has  defined part, yielding the composition $$\tau\scirc \sig: T\scirc S \to A^\perp\vvbar C$$ once we reinstate polarities.  The composition of strategies $\tau\scirc \sig$ is a form of 
synchronised composition of processes followed by the hiding of internal moves,
a view promulgated by Abramsky within traditional game semantics of programs.  

 Two further conditions, {\em receptivity} and {\em innocence},  are demanded of strategies.  The conditions are necessary and sufficient to ensure that copy-cat strategies behave as identities w.r.t.~composition~\cite{lics11}.  Receptivity expresses that any Opponent move allowed from a reachable position of the game is present as a move in  the strategy.    In more detail,   $\sig:S\to A$ is receptive when for any configurations $x$ of $S$ if $\sig x$ extends purely by Opponent events to a configuration $y$ then there is a unique extension of $x$ to a configuration $x'$ of $S$ such that $\sig x' = y$.     
Innocence says a strategy can only adjoin new causal dependencies of the form $\opmove\imc\plmove$, where Player awaits moves of Opponent, 
beyond those already inherited from the game.  

The Scott order makes a reappearance in a technically-important characterisation of strategies $\sig:S\to A$ in a game $A$ as  
certain discrete fibrations from the finite configurations of $S$ and to those of $A$, under the Scott order. (Discrete fibrations correspond to presheaves, a  generalisation of characteristic function to categories, in which `true' and `false' are replaced, respectively, by the set of ways to realise truth and the empty set.) 
The association of a discrete fibration with a strategy respects composition and 
explains how to regard strategies as special forms of profunctor~\cite{fossacs13}. (Profunctors are the generalisation of relations, usually between sets, to categories and have been proposed as a generalised domain theory~\cite{hyland-gendomthy,cattani-winskel}.)
Considering what is lost in the transition from strategies to profunctors   
has led to
an extension of distributed games to
`rooted' factorisation systems, in which strategies may be continuous processes~\cite{
fossacs13}.  

  The literature is often concerned with deterministic strategies, in which Player has at most one consistent response to Opponent.  We can broaden the concept of deterministic strategy to distributed strategies by taking such a strategy to be {\em deterministic}  if consistent moves of Opponent entail consistent moves of Player---see~\cite{lics11,DBLP:journals/fac/Winskel12}.  In general the copy-cat strategy for a game  need not be deterministic.  Copy-cat is however deterministic precisely for games 
which are {\em race-free}, \ie~such that at any configuration, if both a move of Player and a move of Opponent are possible then they may occur together. Deterministic distributed strategies coincide with the {\em receptive} ingenuous strategies of Melli\`es and Mimram~\cite{DBLP:conf/concur/MelliesM07}.

Just as strategies 
generalise relations, deterministic strategies generalise functions. In fact, multirelations and functions are recovered as strategies, respectively deterministic strategies, in the special case where the games are composed solely of Player moves with trivial causal dependency and where only the empty set and singletons are consistent.

As would be hoped the concepts of strategy and deterministic strategy espoused  here reduce to the expected traditional notions on traditional games.  There have also been pleasant surprises.  In the extreme case where games comprise purely Player moves, strategies correspond precisely to the `stable spans' used in giving semantics to nondeterministic dataflow~\cite{SEW}, and in the deterministic subcase one recovers  exactly the {\em stable domain theory} of G\'erard Berry~\cite{berry}.

We have not said much about what it means for a strategy to be winning or optimal,  or how a strategy might be made probabilistic.  This will come.  But first we point out the richness of constructions in the world of distributed  strategies and games.  The language of games and strategies that ensues will be largely stable under the addition of extra features described in the subsequent sections.

 \section{A language for strategies~\cite{lics2014-2}}\label{sec:language}
 
Games  $A, B, C, \cdots$ will play the role of types.  There are operations on games of forming the dual $A^\perp$, 
simple parallel composition $A\vvbar B$,
sum $\Sigma_{i\in I} A_i$ as well as recursively-defined games 
 --- the latter rest on well-established techniques~\cite{icalp82} and will be largely ignored here.  The operation of sum of games is similar to that of simple parallel composition but where now moves in different components are made inconsistent.

Terms denoting strategies 
have typing judgements:
$$
x_1:A_1, \cdots, x_m:A_m \vdash\  t \ \dashv y_1:B_1, \cdots, y_n:B_n\ ,
$$
where all the variables are distinct,
interpreted as a strategy from the game $
 A_1 \vvbar \cdots \vvbar A_m$   to the game $B_1 \vvbar \cdots \vvbar B_n$.
 We can think of the term $t$ as a box with input and output wires for the  variables:
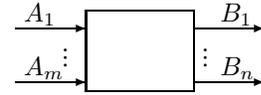
\begin{center}
 \setlength{\unitlength}{2947sp}%
\begingroup\makeatletter\ifx\SetFigFont\undefined%
\gdef\SetFigFont#1#2#3#4#5{%
  \fontfamily{#3}\fontseries{#4}\fontshape{#5}%
  \selectfont}%
\fi\endgroup%
\begin{picture}(2124,699)(889,-598)
\thinlines
{\put(1501,-586){\framebox(900,675){}}
}%
{\put(901,-61){\vector( 1, 0){600}}
}%
{\put(2401,-511){\vector( 1, 0){600}}
}%
{\put(901,-511){\vector( 1, 0){600}}
}%
{\put(2401,-61){\vector( 1, 0){600}}
}%
\put(976,0){\makebox(0,0)[lb]{\smash{{\SetFigFont{9}{14.4}{\rmdefault}{\mddefault}{\updefault}{$A_1$}%
}}}}
\put(976,-451){\makebox(0,0)[lb]{\smash{{\SetFigFont{9}{14.4}{\rmdefault}{\mddefault}{\updefault}{$A_m$}%
}}}}
\put(2626,0){\makebox(0,0)[lb]{\smash{{\SetFigFont{9}{14.4}{\rmdefault}{\mddefault}{\updefault}{$B_1$}%
}}}}
\put(2626,-451){\makebox(0,0)[lb]{\smash{{\SetFigFont{9}{14.4}{\rmdefault}{\mddefault}{\updefault}{$B_n$}%
}}}}
\put(2476,-361){\makebox(0,0)[lb]{\smash{{\SetFigFont{12}{14.4}{\rmdefault}{\mddefault}{\updefault}{$\vdots$}%
}}}}
\put(1301,-361){\makebox(0,0)[lb]{\smash{{\SetFigFont{12}{14.4}{\rmdefault}{\mddefault}{\updefault}{\vdots}%
}}}}
\end{picture}%
\end{center}
 
\noindent
{\bf Duality}  The duality of strategies, that a strategy from $A$ to $B$ can equally well be seen as a strategy from $B^\perp$ to $A^\perp$,  is caught by the rules:
$$
\pfrule{\Gamma, x:A \vdash t \dashv  \Delta}{ \Gamma \vdash t \dashv x:{A}^\perp, \Delta}
 \qquad
\pfrule{\Gamma \vdash t \dashv x:A, \Delta}{ \Gamma, x: {A}^\perp \vdash t \dashv \Delta}
$$

\noindent
{\bf Composition} The composition of   strategies is described in the rule
$$
\pfrule{{\Gamma \vdash t \dashv \Delta \qquad \Delta \vdash u \dashv \Eta} }{\Gamma \vdash 
\stcomp\Delta t u
 \dashv \Eta}
$$
which, in the picture of strategies as boxes, joins the output wires of one strategy to input  wires of the other.

\noindent
{\bf Nondeterministic sum} We can form the nondeterministic sum of strategies of the same type:
$$
\pfrule{\Gamma \vdash t_i \dashv  \Delta \quad i \in I} { \Gamma \vdash \fatbar_{i \in I} t_i \dashv  \Delta}
$$
The construction is like that of the sum of games 
but where the initial Opponent events are identified to maintain receptivity. 
The empty nondeterministic sum 
denotes the minimum strategy in the game $\Ga^\perp\vvbar \Delta$. 

\noindent
 {\bf Pullback} The pullback of a strategy across a map of event structures is itself a strategy~\cite{Probstrats}.  In particular, we 
can form the pullback of two strategies of the same type:
 $$
 \pfrule{\Gamma \vdash t_1 \dashv  \Delta \quad \Gamma \vdash t_2 \dashv  \Delta} { \Gamma \vdash t_1\wedge t_2 \dashv  \Delta}
 $$
Such a strategy acts as the two component strategies agree to act.
 
\noindent
{\bf 
Copy-cat 
terms} 
Copy-cat terms are a powerful way to  lift maps or relations expressed in terms of  maps to strategies. Along with duplication they introduce new ``causal wiring.''
A general copy-cat term takes the form
$$
\pfrule{\Gamma\vdash p':C \qquad \Delta \vdash p:C}
{\Gamma \vdash p \bel_{C} p' \dashv \Delta}\quad p[\emptyset] \bel_C p'[\emptyset]
$$
based on  expressions $p, p'$ for configurations of $C$ 
and their typings; it is stipulated in the formation of a copy-cat term that the configurations denoted by $p$
and $p'$ are in the Scott order initially, when their variables are assigned the empty configuration.  The configurations of the strategy denoted by the copy-cat term are built from 
configurations satisfying the relation $p \bel_{C} p'$ 
by adding  causal links in the manner of copy-cat.
  A term for 
copy-cat arises as a special case, 
$$
x:A \vdash y\bel_A x \dashv y:A \,,
$$
as do terms for the jth injection into and jth projection out of a sum $\Sigma_{i\in I} A_i$ w.r.t.~its component $A_j$,
$$
x:A_j \vdash y\bel_{\Sigma_{i\in I} A_i} j x \dashv y:\Sigma_{i\in I} A_i 
$$
and
$$
x:\Sigma_{i\in I} A_i  \vdash jy\bel_{\Sigma_{i\in I} A_i}  x \dashv y:A_j\,,
$$
as well as terms which split or join `wires' to or from a game $A\vvbar B$.

A map $f:A\to B$ of  games both  lifts to a  strategy $f_!: A\profto B$ described by
$$
x:A\vdash\ y\bel_B f x \ \dashv y:B\,
$$
and  to a strategy $f^*:B\profto A$,
$$
 y:B \vdash\ f x \bel_B y \ \dashv\ x:A\,.
$$
%

Forming the composition $ f^* \scirc t$   pulls back  a 
strategy   
$t$ in $B$ across the map $f:A\to B$ to a strategy in $A$.  It can introduce extra events and dependencies into the strategy.  It subsumes operations of prefixing a strategy by an initial Player or Opponent move.

 \noindent
{\bf  Trace and recursion} 
A {\em trace}, or feedback,  operation is another consequence of such ``wiring.''  
  Given a   strategy $ \Ga,  x:A \vdash  t \dashv y:A,\Delta$  
   represented by the diagram
  
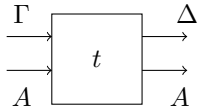
\begin{center}
\begin{tikzpicture}
\node (sigma) at (3, 3) [shape=rectangle, draw, minimum size=1.2cm] { $t$} ;
\draw [->] (1.8, 3.3) -- (2.4, 3.3) ; \node at (2, 3.6) { $\Gamma$ } ;
\draw [->] (3.6, 3.3) -- (4.2, 3.3) ; \node at (4.2, 3.6) { $\Delta$ } ;
\draw [->] (1.8, 2.85) -- (2.4, 2.85) ; \node at (2, 2.5) { $A$ } ;
\draw [->] (3.6, 2.85) -- (4.2, 2.85) ; \node at (4.1, 2.5) { $A$ } ;
\end{tikzpicture}
\end{center}
we obtain 
$$\Ga, \Delta^\perp \vdash t\dashv x:A^\perp, y:A$$ 
which post-composed 
with the term $$x:A^\perp, y:A\vdash  x\bel_A y \dashv\,,$$ denoting the copy-cat strategy $\cc_{A^\perp}$, 
yields 
$$
 \Ga  \vdash \stcomp{x:A^\perp, y:A}{t}{ x\bel_A y}\dashv \Delta\,,
 $$
 representing its trace:  

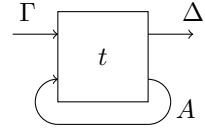
{\centering
\begin{tikzpicture}
\node (sigma) at (3, 3) [shape=rectangle, draw, minimum size=1.2cm] { $t$} ;
\draw [->] (1.8, 3.3) -- (2.4, 3.3) ; \node at (2, 3.6) { $\Gamma$ } ;
\draw [->] (3.6, 3.3) -- (4.2, 3.3) ; \node at (4.2, 3.6) { $\Delta$ } ;

\draw (3.6, 2.7) to [out=0, in=90] (3.9, 2.4) ;
\draw (3.9, 2.4) to [out=270, in=0] (3.6, 2.1) ;
\draw (3.6, 2.1) to (2.4, 2.1) ;
\draw (2.4, 2.1) to [out=180, in=270] (2.1, 2.4) ;
\draw [->] (2.1, 2.4) to [out=90, in=180] (2.4, 2.7) ;
\node at (4.1, 2.3) { $A$ } ;
\end{tikzpicture}
\\}
The composition introduces causal links from the Player moves of $y:A$ to the Opponent moves of $x:A$, and  
 from the Player moves  of $x:A$ to the Opponent moves of $y:A$ --- these are the usual links of copy-cat $\cc_{A^\perp}$ as seen from the left of the turnstyle. This trace coincides with the feedback operation which has been used in the semantics of nondeterministic dataflow (where only
  games comprising solely Player moves are needed)~\cite{SEW}.  

 {\em Recursive definitions} can be achieved from trace with the help of duplication terms, based on a strategy $\delta_A$ from a game $A$ to $A\vvbar A$, roughly, got by joining two copy-cat strategies together: 
%
$$
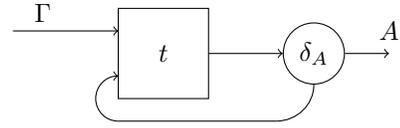
\begin{tikzpicture}
  \node (sigma) at (3, 4) [shape=rectangle,draw, minimum size=1.2cm] {$
t$};
  \draw [->] (1, 4.3) -- (2.4, 4.3) ; \node [above] at (1.4, 4.3) {$\Gamma$} ;
  \node at (6, 4.3) {$A$} ;
  \node (delta) at (5, 4) [shape=circle,draw, minimum size=1] { $\delta_A$ };
  \draw [->] (3.6, 4) to (delta) ; \draw [->] (delta) to (6, 4) ;
  \node (dummy) at (3, 2.5) {} ; 
  \draw (delta.south) to [out=270, in=0] (4.5, 3.1);
  \draw (4.5, 3.1) to (2.4, 3.1);
  \draw (2.4, 3.1) [out=180, in=270] to (2.1, 3.4);
  \draw [->] (2.1, 3.4) to [out=90, in=180] (2.4, 3.7) ;
\end{tikzpicture}
$$

\vskip -0.3in

\noindent
Provided the body $t$ of the recursion respects $\delta_A$ the  diagram above unfolds in the way expected of recursion, to:\\

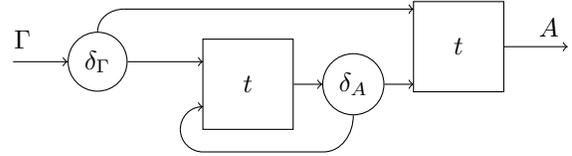
\begin{tikzpicture}
  \node (delta1) at (3, 4.3) [shape=circle,draw] { $\delta_\Gamma$} ;
  \node (input) at (2, 4.3) { } ;
  \node (blabla) at (2, 4.6) { $\Gamma$ } ;
  \node (sigma1) at (5, 4) [shape=rectangle, draw, minimum size=1.2cm] { $t$ } ;
  \node (sigma2) at (7.8, 4.5) [shape=rectangle, draw, minimum size=1.2cm] { $t$ } ;

  \node (delta2) at (6.4, 4) [shape=circle,draw] { $\delta_A$ } ;
  \draw [->] (input.west) -- (delta1.west) ;
  \draw [->] (delta1.east) -- (4.4, 4.3) ;
  \draw [->] (sigma1.east) -- (delta2.west) ;
  \draw [->] (delta2.east) -- (7.2, 4) ;
  \draw (delta1.north) to [out=90, in=180] (3.35, 5) ;
  \draw [->] (3.35, 5) -- (7.2, 5) ;
  \draw (delta2.south) [out=270, in=0] to (6, 3.1) ;
  \draw (6, 3.1) to (4.4, 3.1) ;
  \draw (4.4, 3.1) to [out=180, in=270] (4.1, 3.4) ;
  \draw [->] (4.1, 3.4) to [out=90, in=180] (4.4, 3.7) ;
  \draw [->] (sigma2.east) -- (9.25, 4.5) ;
  \node (bl) at (9, 4.75) { $A$ } ;
\end{tikzpicture} 

In fact, recursive definitions can made more generally, without the use of trace, by exploiting old techniques for defining event structures recursively.  The substructure order $\tri$ on event structures forms a ``large complete partial order,'' continuous operations on which possess least fixed points  --- see~\cite{icalp82,evstrs}.  
Given $x:A, \Ga \vdash t   \dashv y:A$, 
the term $\Ga \vdash  \mu\, x\!:\!A.\, t   \dashv y:A$ denotes the
 $\tri$-least fixed point amongst strategies $X:\Ga\profto A$ of the $\tri$-continuous operation $F(X)= t\scirc (\id_\Ga\vvbar X)\scirc \delta_\Ga$;  here $\sig\tri \sig'$ between two strategies $\sig: S\to \Ga^\perp\vvbar A$ and  $\sig': S'\to \Ga^\perp\vvbar A$ signifies $S\tri S'$ and that the  associated inclusion map $i: S\to S'$ makes $\sig = \sig' i$.

\section{Winning conditions~\cite{lics2012,concur2013}}\label{sec:winconds}
Winning conditions of a game $A$ are given by  specifying a subset   of   {\em winning configurations} $W$. An outcome in $W$ is a win for Player; for simplicity 
assume 
that any other outcome is a win for Opponent.  
A strategy (for Player) is regarded as {\em winning}    if it   always prescribes moves for  Player to  end up in a winning configuration, no matter what the activity or inactivity of Opponent.  Formally, say a configuration is +-maximal if no additional Player moves can occur from it.  Say a strategy  $\sig:S\to A$   is {\em winning}  if  $\sig x$ is in $W$ for all   +-maximal configurations $x$ of $S$.  This can be shown  equivalent to  all plays of $\sig$ against (deterministic) counter-strategies of Opponent result in a win for Player.  

As the dual of a game $A$ with winning conditions $W$ we again reverse the roles of Player and Opponent to get $A^\perp$ and take its  winning conditions to be the set-theoretic complement  of $W$.
In a simple parallel composition of games with winning conditions, $A\vvbar B$,  Player wins if they win in either component.  
With these extensions we can take a winning strategy from a game $A$ to a game $B$, where both games have winning conditions,  to be a winning strategy in the game $A^\perp\vvbar B$.
The choices ensure that the composition of winning strategies is winning.  
In order to have identities w.r.t.~composition, 
we need a condition on games (in fact quite a general condition, implied by race-freeness) to guarantee that copy-cat strategies are winning. 

Often questions can be reformulated in terms of the existence of a winning strategy for Player or Opponent.  For this it is important to know whether there is a winning strategy for one of the players --- the issue of {\em determinacy} of the game.  A famous result is  Martin's theorem, at the boundary of the power of set theory, which shows that sequential games are determined if their winning conditions form a Borel set~\cite{MartinBorel}.  
Distributed games with races need not be determined even though their winning configurations form a Borel set.  However distributed games are determined provided that they are race-free and satisfy a structural condition, ``bounded concurrency''~\cite{concur2013}. Bounded concurrency expresses that no move of one of the players can be concurrent with infinitely many moves of the other --- a condition trivially satisfied when \eg~all plays are finite, the games are sequential, or the games have rounds where simple choices are made (usual in traditional accounts).   Like race-freeness, bounded concurrency is necessary for the Borel determinacy of distributed games.  

\section{Probabilistic event structures and strategies~\cite{Probstrats}}\label{sec:prob}
The extension of 
distributed strategies to  probabilistic strategies required a new general definition of probabilistic event structure.   
A probabilistic event structure essentially comprises an event structure together with a  continuous valuation on the Scott-open sets of its domain of configurations.\footnote{A {\em Scott-open} subset of configurations is upwards-closed w.r.t.~inclusion and such that if it contains the union of a directed subset $S$ of configurations then it contains an element of $S$. 
A {\em continuous valuation} is a function $w$ from  the Scott-open subsets of  $\iconf E$ to $[0,1]$ which is
 {\em (normalized)} \  $w(\iconf E) = 1$;    {\em (strict) }\  $w(\emptyset) = 0 $;
 {\em  (monotone)} \ $U \subseteq V \implies w(U)\leq w(V)$;
 \noindent{\em  (modular)} \ $w(U \cup V) + w(U\cap V) = w(U) + w(V)$; and
 \noindent {\em  (continuous) }\ $w(\bigcup_{i\in I} U_i) = {\rm sup}_{i\in I} w(U_i)$, 
 for {\em directed} unions. The idea:  $w(U)$ is the probability of a result in open set $U$.
 } The continuous valuation assigns a probability to each open set and can then be extended to a probability measure on the Borel sets~\cite{jonesplotkin}.  However open sets are several levels removed from the events of an event structure, and an equivalent but more workable definition is obtained by considering the probabilities of basic open sets, generated by single finite configurations;  for each finite configuration this specifies the probability of obtaining a result which extends the finite configuration.  Such  valuations on configurations determine the continuous valuations from which they arise,  and can be characterised through the device of ``drop functions'' which measure the drop in probability across certain generalised intervals.  The characterisation yields a workable general definition of probabilistic event structure as event structures with {\em configuration-valuations}, \viz~functions from finite configurations to the unit interval for which the drop functions are always nonnegative.

In 
detail, a {\em probabilistic event structure} comprises an event structure $E$ with  a {\em configuration-valuation}, a function  $v$ from the finite configurations of $E$ to the unit interval  which  is 
 \begin{itemize}
\item[]
{\em (normalized)}\  $v(\emptyset) =1$   and has 
\item[] 
{\em (non\,$-$ve drop)}\ 
  $\drp y{x_1, \cdots,x_n}  \geq 0$  when $y\subseteq x_1, \cdots,x_n$ for finite configurations $y, x_1, \cdots,x_n$ of  $E$, 
\end{itemize}
where the ``drop'' across the generalized interval starting at $y$ and ending at one of the  $x_1, \cdots,x_n$ is given by 
 %
$$ 
\drp  y{x_1, \cdots,x_n}  \eqdef v(y) - \sum_I (-1)^{|I|+1} v(\bigcup_{i\in I} x_i) 
$$
---the index $I$ ranges over nonempty $I \subseteq \setof{1,\cdots, n}$ such that the union $\bigcup_{i\in I} x_i$ is a configuration. The ``drop''  $\drp  y{x_1, \cdots,x_n}$ gives the probability of the result being a configuration which includes the configuration $y$ and does not include any of the configurations $x_1, \cdots,x_n$.\footnote{Samy Abbes has pointed out that the same ``drop condition" appears in early work of the Russian mathematician V.A.Rohlin~\cite{rohlin}(as relation (6) of Section 3, p.7). Its rediscovery  in the context of event structures was motivated by the need to tie probability to the occurrences of events;  
it is sufficient to check the `drop condition' for $y\longcov{} x_1, \cdots,x_n$, in which the configurations $x_i$ extend $y$ with a single event.}

This prepares the ground for a general definition of distributed probabilistic strategies, based on event structures.  One  hurdle is that in a strategy it is impossible to know the probabilities assigned by Opponent.  A probabilistic strategy  in a game $A$, presented as a race-free \eswp,  is a   strategy $\sig:S\to A$ in which we
  endow $S$ with probability, while 
 taking account of the fact that in a strategy Player can't be aware of the probabilities assigned by Opponent.  
We do this by extending the notion of configuration-valuation so that: 
causal independence between Player and Opponent moves entails their probabilistic independence,  or equivalently, so probabilistic dependence of Player on Opponent moves will presuppose their causal dependence (the effect of the condition of ``$\pm$-independence'' below); the ``drop condition'' only applies to moves of Player.  Precisely,
a  {\em configuration-valuation} is now a function $v$, from finite configurations of $S$  to the unit interval, which is
 \begin{itemize}
\item[]{\em (normalized)}\  
$
 v(\emptyset ) =1$, has 
\item[]
{\em ($\pm$-independence)}\  
 $v(x) = v(y)$ when  $x\subseteq^- y$ for finite configurations $x$, $y$ of 
  $S$,
 and satisfies 
  the 
\item[]{\em (+ve drop condition)}\ 
 $
 \drp y{x_1, \cdots, x_n} \geq 0
 $
 when  $y\subseteq^{+} x_1, \cdots, x_n$ for finite configurations of $S$.  
  \end{itemize}
We return to the point that ``$\pm$-independence'' expresses that causal independence between Player and Opponent moves entails their probabilistic independence.  Consider two moves, $\plmove$ of Player and $\opmove$ of Opponent  able to occur independently, \ie~concurrently, at some finite configuration $x$, taking it to the configuration $x\cup\setof{\plmove,\opmove}$.  There are intermediate configurations $x\cup\setof{\plmove}$ and $x\cup\setof{\opmove}$ associated with just one additional move. The condition of ``$\pm$-independence''  
ensures
$v(x\cup\setof{\plmove,\opmove}) = v(x\cup\setof{\plmove})$, \ie~the probability of $\plmove$ with $\opmove$ is the same as the probability of $\plmove$ at configuration $x$.  At  $x$ 
the probability of the Player move  conditional on the Opponent move equals the probability of the Player move --- the moves are probabilistically independent.

Probabilistic strategies  compose --- in the proof `drop functions' come into their own --- with probabilistic copy-cat strategies as identities because we restrict to race-free games.  The result of a play between Player and Opponent in a game will be a probabilistic event structure.\footnote{The use of  ``schedulers to resolve the probability or nondeterminism'' in earlier work is subsumed by that of probabilistic and deterministic counter-strategies. Deterministic strategies coincide with those with assignment one to each finite configuration.}   

\section{Quantum event structures and strategies~\cite{Probstrats, prakash}}\label{sec:quan}
In a {\em quantum event structure}  the events of an event structure are interpreted as unitary or projection operators in a 
Hilbert space in such a way that concurrent events are interpreted by commuting operators.  
 Unitary operators are associated with events of preparation, such as a change of coordinates with which to make a measurement or a time period over which the system is allowed to evolve undisturbed.  Projection operators are associated with events of elementary tests.    A configuration of the event structure is thought of as a distributed quantum experiment; it describes which events of preparation and tests to perform and their partial order of dependency.  The whole event structure can be thought of as a nondeterministic quantum experiment,  and gives us the extra latitude to define a probabilistic quantum experiment as quantum event structure which also carries the structure of a probabilistic event structure.  A quantum event structure comes with an initial state in the form of a density operator.

In more detail, a {\em quantum event structure} (over a Hilbert space with countable basis) comprises an initial state  
 given by a density operator $\rho$  and an event structure $(E, \leq, \Con)$ with an assignment  $Q_e$ of projection or unitary operators   to events $e$ of $E$ such that if two events $e_1$ and $e_2$ are concurrent then the operators $Q_{e_1}$ and $Q_{e_2}$ commute.  
Each finite configuration, $x$ of $E$, determines an operator $A_x$ got as   the composition 
$$Q_{e_n} Q_{e_{n-1}} \cdots Q_{e_2} Q_{e_1}$$ for some 
serialization 
$e_1, e_2, \cdots, e_n$ of $x$ in which $\setof{e_1,   \cdots, e_i}$ is a configuration for all $i\leq n$.  
The operator is well-defined as any two serializations of $x$ are obtainable, one from the other, by successively interchanging concurrent events.
A quantum event structure assigns an intrinsic weight   
$v(x)\eqdef \Tr(A_x^\dagger A_x \rho)$, the trace of the operator $A_x^\dagger A_x \rho$,  to each finite configuration $x$; in the case where $\rho$ corresponds to a pure state $\ket \psi$ the weight will be the square of the norm of $A_x\ket\psi$.  
This does not make the {\em whole} event structure into a 
probabilistic event structure, but it does do so locally: for any configuration 
$w$ of  $E$,  the function $v$ is a configuration-valuation on the event structure obtained by restricting $E$ to the   events $w$.

Quantum theory is often described as a contextual theory,  in that it is only sensible to consider outcomes w.r.t.~a specified measurement context~\cite{samson-adam}.  In a quantum event structure configurations assume the role of measurement contexts;
w.r.t.~a measurement context expressed as a configuration,  the sub-configurations  constitute the possible outcomes. This gives a non-traditional take on the consistent-histories approach to quantum theory, which provides  decoherence 
conditions on histories to pick out those subfamilies of  histories over which it is meaningful to place a probability distribution; the approach via quantum event structures bypasses the  decoherence
 conditions usually invoked~\cite{Griffiths}. 

A {\em quantum strategy} is taken to be a distributed probabilistic strategy on a game  which also carries the structure of a quantum event structure, so that moves perform operators on a Hilbert space.
In a quantum game Player and Opponent interact to jointly create a  probabilistic distributed experiment on a quantum system~\cite{grabbe2005introduction}.   Extensions with pay-off and to games of imperfect information are dealt with as below.

\section{Payoff~\cite{payoff}}\label{sec:payoff}
  We can add  {\em payoff} to a game $A$ as a function $X$ from configurations to the real numbers.  We can extend the operations of dual and simple parallel composition of games  to games with payoff.  As the payoff $X_{A^\perp}$ of the dual game $A^\perp$ we take the negative of the payoff $X_A$ of $A$, 
  $
  X_{A^\perp}(x) =  - X_A(x)
  $
  on configurations $x$ --- this implicitly makes the games zero-sum. 
  As the payoff of $A\vvbar B$ we take sums of the payoffs,
  $
  X_{A\vvbar B}(x) = X_A(x_1) + X_B(x_2)
  $
  on  a configuration $x$ of $A\vvbar B$, with components $x_1$ in $A$ and $x_2$ in $B$.  
  
For such quantitative games, determinacy is expressed in terms of the game possessing a  {\em value},  a form of minimax property.   The interest is now focussed on {\em optimal} strategies which achieve the value of the game.  
How we proceed to associate a value with a distributed game slightly differs according to whether  strategies are assumed probabilistic.  
 
In the probabilistic case, assume that payoff function $X$ on the game $A$  is  Borel measurable.  
Given a probabilistic strategy in $A$, so a strategy $\sig:S\to A$ and a configuration-valuation $v_S$ for $S$,
and a probabilistic 
  counter-strategy in $A$, so a strategy  $\tau:T\to A^\perp$ with a configuration-valuation $v_T$ for $T$, we obtain their composition before hiding as the pullback 
  
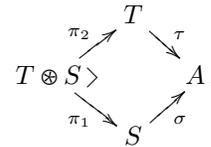
\[
\xymatrix@R=10pt@C=10pt{
&  T	\ar[dr]^{  \tau}\\
\!\!\! T\sscirc S	 \ar[ur]^{\pi_2}
		\ar[dr]_{\pi_1}
		\ar@{}[rr]|{ \ \ }
		\pb{0}&&
A \,.\\
 &S 	\ar[ur]_{\sigma }
}
\]
  The pullback is associated with the map  $f\eqdef \sig\pi_1 =\tau\pi_2:T\sscirc S\to A$.  The event structure $T\sscirc S$ comes equipped with a configuration-valuation $v(x) =v_S(\pi_1 x) \times v_T(\pi_2 x)$ on its finite configurations $x$.  This determines a Borel probability measure $\mu_{v}$   on all the configurations of $T\sscirc S$.  
     The {\em expected payoff} is obtained as the Lebesgue integral $${\bf E}_{\sig,\tau}(X) = \int X (f x) \ d\mu_{v}(x)\,$$ 
  over the configurations $x$ of $T\sscirc S$.  Define 
  $$v(A) \eqdef \Sup_{\sig:A}\Inf_{\tau:A^\perp} {\bf E}_{\sig,\tau}(X)\,,$$
where   $\sigma : A$ signifies a probabilistic strategy in $A$  and $\tau:A^\perp$ a probabilistic strategy counter-strategy.  
The game is {\em determined} w.r.t.~probabilistic strategies if $v(A) =-v(A^\perp)$, with $v(A)$ then being called the {\em value} of the game; since the order of $\Sup$ and $\Inf$ are reversed in $v(A)$ and $-v(A^\perp)$, determinacy amounts to a minimax property.  For a determined game, an {\em optimal} strategy is one $\sig:A$ for which 
$$v(A)=\Inf_{\tau:A^\perp} {\bf E}_{\sig,\tau}(X)\,.$$
After~\cite{concur2013}, it is reasonable to conjecture that any bounded concurrent, race-free distributed game with measurable payoff function  is determined w.r.t.~probabilistic strategies --- though this has not yet been proved.  
 
Without probability, for distributed strategies which are in general nondeterministic, we instead consider both the {\em optimistic} $v^\uparrow(A)$ and {\em pessimistic} value $v^\downarrow(A)$   of a game $A$. The optimistic value  stems from taking the result of the interaction of a strategy   and a counter-strategy to be the supremum over the values resulting from all maximal plays (as the strategies are nondeterministic there may be many maximal plays).  The  pessimistic value is defined analogously using infimum.  We
  say a game is determined with value $v(A)$ if 
$$
v(A)=v^\uparrow(A) = v^\downarrow(A) = -v^\downarrow(A^\perp) = -v^\uparrow(A^\perp)\,.
$$
In this case an optimal strategy is one with pessimistic value that of the game.  
It is to be expected that  
 any bounded concurrent, race-free distributed game with Borel measurable payoff function is determined, 
 though this has so far only been proved for well-founded games, in which all configurations are finite~\cite{payoff}.  In {\it op. cit.}~it is shown that optimal strategies compose, so fit within the language of Section~\ref{sec:language}.    This 
makes a start on a 
 general
 {\em structural game theory}  in which games and optimal strategies are built up compositionally.  

\section{Imperfect information~\cite{Dexter,Probstrats}}\label{sec:impinfo}

As they stand the games so far described are  
games of {\em perfect information}.  In  games of  {\em imperfect information}  
 some moves are masked, or inaccessible,  
 and strategies with dependencies on unseen moves are ruled out.  It is straightforward 
 to extend distributed games to games with imperfect information  in way that respects the operations of distributed  games and strategies~\cite{Dexter} and does not disturb the addition of extra features such as probability. A fixed preorder of {\em levels}
  $(\Lambda, \preceq)$ is pre-supposed.  The levels are to be thought of as levels of access,  or permission.   Moves in games and strategies are to respect levels:  a move is only permitted to  causally depend on moves at equal or lower levels; it is as if from a level only  moves of equal or lower level can be seen. 
  A  distributed game of imperfect information,  a {\em $\Lambda$-game}, comprises a game $A$  with a {\em level function} $l:A\to \Lambda$ such that if
$a\leq_A a' $ then 
$ l(a)\preceq l(a')
$ 
 for all moves $a, a'$ in $A$.  A  {\em $\Lambda$-
 strategy} in the $\Lambda$-game is a   strategy $\sig:S\to A$ for which 
if $
s\leq_S s' $
then 
$l\sig(s) \preceq l\sig(s')
$ 
  for all $s, s'$ in $S$.  
   One interpretation of 
 $\Lambda$, pertinent to the treatment of quantum strategies, is as space-time with $\lam \preceq\lam'$ meaning there is a causal curve from $\lam$ to $\lam'$.

 In particular, {\em Blackwell games}~\cite{det-blackwell}, of central importance in logic and computer science,  become a special case of  probabilistic $\Lambda$-games with payoff.  Blackwell games are  games of imperfect information for which an appropriate choice of $\Lambda$ is the infinite  event structure:
$$
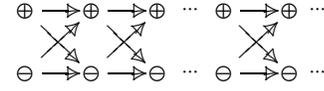
\xymatrix@R=12pt@C=12pt{
\plmove\ar@{|>}[r]\ar@{|>}[rd]&\plmove\ar@{|>}[r]\ar@{|>}[rd]&\plmove\ar@{}[r]|{\cdots}&
\plmove\ar@{|>}[r]\ar@{|>}[rd]&\plmove \ar@{}[r]|{\cdots}&
\\
\opmove\ar@{|>}[r]\ar@{|>}[ru]&\opmove\ar@{|>}[r]\ar@{|>}[ru]&\opmove\ar@{}[r]|{\cdots} &
\opmove\ar@{|>}[r]\ar@{|>}[ru]&\opmove\ar@{}[r]|{\cdots} &
}
$$
A Blackwell game is given by  $A$, a race-free distributed game with payoff $X$,  for which there is 
a (necessarily unique) polarity-preserving rigid map from $A$  to $\Lambda$---this map becomes the 
  level function. Moves in $A$ occur in rounds comprising a choice of move for Opponent and a choice of move for Player made independently. 
  Traditionally, in Blackwell games 
a strategy (for Player) is a `total' $\Lambda$-strategy in such a $\Lambda$-game --- strategies are restricted to those  assigning   
probability distributions at each round. In fact, 
the existing literature is most often concerned with  {\em total} 
strategies which always progress, which we can express very generally for distributed strategies
by insisting there is zero probability of ending at a configuration from which a Player move is possible. 
 
Based on our experience and the insights of Martin~\cite{det-blackwell}, the strongest determinacy result we can hope for  in distributed games with imperfect information is the conjecture that  any finite-width, race-free distributed game with imperfect information, and Borel measurable payoff function  is determined w.r.t.~probabilistic strategies. This result would go a considerable way to subsuming the theory of Blackwell games within distributed games.

\section{Games with symmetry~\cite{CCW}}\label{sec:sym}
  There are several reasons to consider symmetry in games, situations where distinct plays are essentially similar to one another.  Symmetry can help in the analysis of games, by for instance reducing the number of cases to consider.  Symmetry   can also help compensate for the overly-concrete nature of event structures in representing games: many useful operations on games which are not algebraic operations 
become so {\em up to symmetry}. 
In particular, through symmetry we can support 
operations on types in the form of {\em monads} --- well-known from functional programming ---
  and this leads to richer type systems and, in particular, the ability to support backtracking in games.  Although backtracking is not allowed in many games it is important in logic, in representing assertions by games and proofs as strategies between them, when it is very common for the proof  to use the same assertion several times.  Similarly, in computer science, where  types are often represented by games and programs by strategies, a program may generally reuse its input.  Backtracking is supported through 
a monad $!A$ on a game $A$ where $!A$ consists of many copies of moves in the original game $A$; while copies of the same move are distinct from each other they are essentially similar, a similarity we can express through symmetry in games.  

The treatment of 
symmetry in games stems from earlier work on symmetry in event structures~\cite{ESS} and makes use of a general method of open maps for defining the equivalence of bisimulation in a variety of models~\cite{JNW}. Briefly, a symmetry in an event structure $E$ is expressed as a bisimulation equivalence, given as a span of open maps $l, r:\sym E \to E$.  The finite configurations of $\sym E$ correspond to bijections between finite configurations of $E$ expressing that one configuration is similar to the other.  Together the finite configurations of $\sym E$ correspond to an {\em isomorphism family} comprising a non-empty family of bijections  
$
\theta : x \cong_E y 
$
between pairs of finite configurations of $E$ such that: 

\noindent
(i) for all identities on finite configurations $\id_x: x \cong_E x$; if $\theta : x \cong_E y$, then 
the inverse $\theta^{-1}:y\cong_E x$; and 
if $\theta : x \cong_E y$ and $\varphi:y\cong_E z$, then 
their composition
$\varphi\theta:x\cong_E z$.

\noindent
(ii) for $\theta : x \cong_E y$ 
and finite configurations $x'\subseteq x$ 
there is  a (necessarily unique)  finite configuration $y'$ of $E$ with $y'\subseteq y$ such that the restriction 
$\theta' : x'\cong_E y'$. 
 
 \noindent
(iii) for $\theta : x \cong_E y$ 
whenever $x\subseteq x'$ with $x'$ a finite configuration, 
 there is an extension of $\theta$ to  $\theta'$ so
 $\theta' : x'\cong_E y'$ 
 for some (not necessarily unique)  configuration $y'$ with $y\subseteq y'$.

\noindent
(Because of (ii)  the bijections in the isomorphism family respect the partial order of causal dependency on configurations inherited from $E$.)  

A total map $f:A\to B$ between \essws~{\em preserves  symmetry} when $x \stackrel{\theta}{\iso_A} y$ implies  $fx \stackrel{\sym f\theta}{\iso_B} fy$, where $\sym f\theta$ is the composite bijection $fx\iso x\stackrel{\theta}{\iso_A} y\iso fy$.  While two  total maps $f,g:A\to B$ preserving symmetry, 
are {\em equivalent up to symmetry}, written $f\sim g$,  if $fx\stackrel{\phi_x}{\iso_B} gx$ for all finite configurations $x$ of $A$, where $\phi_x$ is the composite bijection $fx\iso x\iso gx$.  

With the addition of symmetry event structures no longer have pullbacks.  But they do have pseudo pullbacks, a form of pullback up to symmetry.  The {\em pseudo pullback} of total 
maps $f:A\to C$, $g:B\to C$   comprises 
$$
\xymatrix@R=8pt@C=8pt{
&A\ar[rd]^{f}&\\
P
\ar[ur]^{\pi_1}\ar[rd]_{\pi_2}&\sim&C\\
&B\ar[ur]_{g}&}
$$
where 
$f \pi_1 \sim g \pi_2 $ with   the further property that 
for any object 
$D$ and maps $p_1:D\to A$ and $p_2:D\to B$ such that $f  p_1 \sim g p_2 $, there is a unique map $h:D\to P$ such that $p_1 = \pi_1 h$ and $p_2 = \pi_2  h$.  The pseudo pullback is defined up to isomorphism. 
Concretely, finite configurations of $P$ correspond to the composite bijections 
$$
\theta:x\iso fx  \stackrel{\varphi}{\iso_C}  gy\iso  y
$$
 between finite configurations $x$ of $A$ and $y$ of $B$ such that $f x\stackrel{\varphi}{\iso_C} g y$
for which the transitive relation generated on $\theta$ by  $(a,b)\leq (a',b')$ if $a\leq_A a'$ or $b\leq_B b'$ 
  forms a partial order.
  
A particular pseudo pullback is obtained as
$$\xymatrix@R=8pt@C=8pt{
&A\ar[dr]^{\id_A}&\\
\sym A\ar[ur]^{l_A}\ar[dr]_{r_A}&\sim&A\\
&A\ar[ur]_{\id_A}&}
$$
It recovers $\sym A$, the symmetry on $A$, but as an \esws~so {\em  itself equipped with symmetry} $\ssym A$.  This pseudo pullback is that associated with a {\em path object} of homotopy generally written $A^I$, where $I$ stands for (a generalization of) the unit interval: asserting $\theta:x\iso_A y$, that a bijection between two configurations is in the isomorphism family of $A$, is analogous  to specifying a path from $x$ to $y$.  There are also {\em cylinder objects}. In fact \essws~have the structure of a homotopy category.   Just as in homotopy theory one considers operations up to homotopy it is sensible to consider operations within event structures {\em up to symmetry}.  

A {\em game with symmetry} is represented by 
an event structure with polarity and symmetry (henceforth an e.p.s.).
The operations of 
dual and simple parallel composition of games extend to games with symmetry.  The symmetry of $A^\perp$ coincides with that of $A$, while the symmetry of $A\vvbar B$ is built as $\sym A\vvbar \sym B$.   

A strategy in a game with symmetry $A$ will be a total map $\sig:S\to A$ of \essws.
We regard two strategies $\sigma:S\to A $ and $\sigma':S'\to A$  as
{\em equivalent} when there are maps $f:S\to S'$  and $g:S'\to S$ such that $\sig= \sig' f$ and $\sig'= \sig g$
with
$gf\sim \id_S$ and $fg\sim \id_{S'}$. 
Following the now familiar pattern a strategy between games with symmetry, from $A$ to $B$, will be a strategy in $A^\perp\vvbar B$. 

The addition of polarity and symmetry brings a new richness to the configurations of an event structure. 
The Scott order becomes a {\em Scott category},     where now maps between configurations are obtained as compositions  
$$
y \supseteq ^- y'  \stackrel{\theta}{\iso_A} x' \subseteq^+ x\,.
  $$
 
 This fact drives the definition of copy-cat strategy for a game with symmetry $A$.  The configurations of  $\CC_A$,  in copy-cat $\cc_A:\CC_A\to A^\perp\vvbar A$, correspond to instances of maps in the Scott category, to those configurations $x$ of $A^\perp\vvbar A$ for which
 $$
 x_2 \supseteq^- x_2'  \stackrel{\theta}{\iso_A} x_1' \subseteq^+ x_1\,,
 $$
 for some instance $x_2'  \stackrel{\theta}{\iso_A} x_1'$  of the  isomorphism family of $A$.  The  symmetry in copy-cat is constructed in a similar way, as
 $$
 \sym{\CC_A} = \CC_{\sym A}\,.
 $$
 The definition of the composition $\tau\scirc \sig: T\scirc S \to A^\perp\vvbar C$ of strategies $\sig:S\to A^\perp\vvbar B$ and $\tau:T\to B^\perp \vvbar C$ between games with symmetry is exactly analogous to that without symmetry but using pseudo pullbacks in place of pullbacks.  The symmetry on 
  $T\sscirc S$, composition before hiding, satisfies
  $$
  \sym{T\sscirc S}= \sym{T}\sscirc \sym S\,.
  $$
The symmetry on $T\scirc S$, composition after hiding, is a little more complicated than $\sym{T}\scirc \sym S$  but only because the latter fails to satisfy joint monicity required of an equivalence.  (For this and other reasons it would be an interesting exercise to redevelop symmetry on games by relaxing the present definition, in which a symmetry is expressed as a bisimulation equivalence, to one in which the requirement of joint monicity is dropped.) 
  
  Again we can ask for conditions on strategies that exactly ensure that copy-cat behaves as identity, more precisely that composition with copy-cat yields an equivalent strategy.
    The necessary and sufficient conditions are strengthenings of receptivity and innocence, seen earlier,  to take account of symmetry:
  in addition, the strategy should be
   {\em  strong-receptive}, \ie~receptivity should also respect symmetry, and 
 {\em saturated}, \ie~behave identically over symmetric parts of the game --- see~\cite{CCW} for the precise definitions.
    
The language of strategies extends to games with symmetry (pseudo pullbacks replace pullbacks) and allows extra types and terms because now in the presence of symmetry there are many $\sim$-monads (technically pseudo monads that satisfy the monad laws up to symmetry).  One simple example is the $\sim$-monad constructing $!A$, 
infinitely many copies of a game $A$,  one for each natural number,  all made similar to each other through the appropriate symmetry.  The unit of the $\sim$-monad essentially injects $A$ into one chosen copy --- which one does not matter, up to symmetry --- and the multiplication essentially flattens the ``array'' of copies, one for each pair of natural numbers, to the ``row'' $!A$.  This $\sim$-monad is central to AJM games~\cite{AJM}.  
 
Under fairly general conditions $\sim$-monads on \essws~lift to monads on $\sim$-strategies, expressible within a slight extension of the language of strategies.  Given such a monad $T$ its unit and multiplication lift to strategies via  copy-cat terms and 
the action of $T$ itself on a strategy is essentially to enclose it within a ``$T$-box'' --- an additional construction in the language. 
 Because of the duality of strategies, a $\sim$-monad on strategies can also be viewed as a $\sim$-comonad.  In particular there are comonads to allow back-tracking of the form required in traditional   AJM and HO games~\cite{AJM,HO,CCW}.  

\section{Extensions and limitations}\label{sec:extnsandlims}
The story so far has been quite well-rounded, I believe.  In this section I'll try to gather the loose ends of which we're presently aware.  In particular, I'll try to give an idea of an obstacle on which we are currently working, the extension of  strategies with ``parallel causes."

First though, note that the language for strategies simultaneously supports both a mathematical  and an operational semantics~\cite{lics2014-2}. The operational semantics, presented as rules for transitions, gives a view of strategies as processes.  It comes at the cost of taking internal, previously hidden, 
events seriously.  Once one does so, questions as to what strategies may and must do arise.  Fortunately it can be shown that as regards the may-and-must behaviour of strategies one can circumvent internal events and in their place work with a notion of maximal configuration (called ``stopping configuration'') divorced from order-theoretically maximal configurations.  In Section~\ref{sec:winconds} +-maximal configurations played a key role; 
their role can be taken over straightforwardly by stopping configurations.

One limitation that is not seen when working with purely nondeterministic strategies has revealed itself when strategies are made probabilistic.  The simple event structures of Section~\ref{sec:evstrs} on which we have based games and strategies do not support ``parallel causes" and this has the consequence that certain informal but intuitively convincing strategies are not expressible.  

Probabilistic strategies, as presented,  do not  cope with stochastic behaviour, \eg~races as in the game 
$$\small
\xymatrix@R=14pt@C=13pt{\opmove\ar@{~}[r]&\plmove}.$$  To do such we would expect to have to equip events in the strategy with stochastic rates.  So this is to be expected --- and isn't  hard to do if internal events are not hidden. But at present probabilistic strategies do not cope with benign Player-Player races either!  Consider the game
$$
\small
\xymatrix@R=4pt@C=0pt{
&\plmove&\\
 \opmove& &\opmove}
$$
where Player wins if a play of any $\opmove$ is accompanied by the play of $\plmove$ and {\it vice versa}.  Intuitively a winning strategy would be got by  assigning watchers (in the team Player) for each $\opmove$  who on seeing their $\opmove$ race to play $\plmove$.  This strategy should win with certainty against any counter-strategy: no matter how Opponent plays one or both of their moves at least one of the watchers will report this with the Player move.   But we cannot express this with prime event structures.  The best we can do is a probabilistic strategy
$$
\small
\xymatrix@R=14pt@C=5pt{
\plmove\ar@{~}[rr]&&\plmove\\
\ar@{|>}[u] \opmove& &\ar@{|>}[u]\opmove}
$$
with configuration-valuation assigning 1/2 to configurations containing either Player move and 1 otherwise.  
Against a counter-strategy with Opponent playing one of their two moves with probability  1/2 this strategy 
only wins half the time.  In fact,  the strategy together with the counter-strategy form a Nash equilibrium when a winning configuration for Player is assigned payoff +1 and a loss $-1$.
This strategy really is the best we can do presently in that it is optimal amongst those expressible using the simple event structures of Section~\ref{sec:evstrs}.  

If we are to be able to express the intuitive strategy which wins with certainty
we need to develop distributed probabilistic strategies to allow such parallel causes  as in  `general event structures' $(E, \vdash, \Con)$ which allow \eg~two distinct compatible causes $X\vdash e $ and $Y\vdash e$ (see~\cite{evstrs}).  In this specific strategy both Opponent  moves would individually enable the Player move, with all events being consistent.
But it can be shown that general event structures do not support an appropriate operation of hiding.  Nor is it clear how within general event structures one could express a variant of the strategy 
 in which the two watchers succeed in reporting the Player move with different probabilities. 

We are currently working on two ways to extend probabilistic strategies with parallel causes in such a way that they still support an appropriate operation of hiding; we hope and expect the two ways will converge.  One way is roughly to extend general event structures so that consistency is between different enablings rather than events; an example shows an extension of this nature is needed if it is to support hiding.  The other way 
 involves a monad $?$ up to symmetry constructed so strategies $\sig:S\to ?A$ can express several parallel ways in which a Player move can be enabled.

\section{The future}\label{sec:future}

The over-abstraction of traditional denotational semantics has largely removed it from algorithmic concerns.  This has led to an artificial division between the semantics of programming languages (concerning their formalisation, algebra and proof of correctness) and algorithmics (concerning the efficiency and classification of computational problems), 
whereas in reality both have need of techniques of the other.  
The division reflects a fundamental tension between the approaches of 
semantics and algorithmics.  
Whereas semantics seeks denotations and equivalences sensible within a very broad range of contexts (generally that allowed by the programming languages of interest) algorithmics is often concerned with optimal methods within very specialised contexts (often algorithms to perform some specific task).
The seemingly-conflicting approaches of semantics and algorithmics can be reconciled provided a semantics can support contextual reasoning and is refined enough to express algorithmic concerns.

Distributed games give a faithful operational account of algorithms while having a rich algebraic structure. 
As one example, a basic tool for the efficient verification of systems is that of binary decision diagrams (bdds); they
are precisely sequential strategies between games representing the booleans.   
Another example is that
important equivalences between programs can sometimes be settled automatically by exploiting the game semantics of programs, together with automata 
representations of strategies~\cite{ong}.
At a fundamental level, deep questions in complexity theory and logic can often be reformulated in terms of games.  
%
The idea of a strategy {\em from} one game {\em to} another was used by Conway in developing his ``surreal numbers'' as strengths of games~\cite{conway}. Once one has a winning/optimal strategy from a game $A$ to a game $B$ by pre-composition it provides a way to convert a winning/optimal  strategy of $A$ to a winning/optimal strategy of $B$; in this way it reduces finding a winning strategy of $B$ to the problem of finding one for $A$.  Similar  reductions are plentiful in algorithmics and logic.  Perhaps they could be systematized through distributed games, which can potentially support a range of reductions in the form of strategies from a game $!A$ to a game $?A$ w.r.t~monads $!$ and $?$ up to symmetry.  
At the very least the determinacy result of Section~\ref{sec:impinfo} would provide a new facility in expressing computational problems in terms of the existence of a strategy
--- presently, often problems are artificially forced into a formulation using Blackwell games. 
Nor at a general level should we underestimate the power of games in providing a vocabulary common to both semantics and algorithmics.

The fundamental nature and range of distributed games and strategies means that they have the  potential to make revolutionary changes to the way we think about computing systems, and through these, other subjects too.  But this will not come easily.   It requires 
significant demonstration of their use in programming, verification, logic
and,   most importantly, in the beneficial cross-over between the algebraic and algorithmic worlds that they simultaneously inhabit.  

\subsection*{Acknowledgements}
This paper is dedicated to Pierre-Louis Curien on his 60th birthday.  It's a great pleasure for me to acknowledge Pierre-Louis's inspiration and friendship over the years,   inspiration and friendship  that go back to our thesis work and first meetings in Sophia Antipolis and Paris in 1979. 

Thanks to Samy Abbes, Nathan Bowler, Simon Castellan, Pierre Clairambault, Pierre-Louis Curien, Marcelo Fiore,  Mai Gehrke, Julian Gutierrez, Jonathan Hayman, Martin Hyland, Alex Katovsky, Marc Lasson, Paul-Andr\'e Melli\`es, Samuel Mimram, Gordon Plotkin, Silvain Rideau and Sam Staton for helpful discussions.  The support of  
Advanced Grant ECSYM of the  European Research Council is acknowledged with gratitude.  


\bibliographystyle{IEEEtran} 

\bibliography{biblio}


\end{document}